\documentclass[amsmath,aps,showpacs,preprint]{revtex4}
\begin{document}

\title{ Anomalous scaling of conductance cumulants in one dimensional
Anderson localization}
\author{Jean Heinrichs}
\email{J.Heinrichs@ulg.ac.be}
\affiliation{Institut de Physique, B{5}, Universit\'{e} de Li\`{e}ge, Sart
Tilman, B-{4000} Li\`{e}ge, Belgium}
\date{\today}

\begin{abstract}
The mean and the variance of the logarithm of the conductance ($ln g$) in the localized regime in the one-dimensional Anderson model are calculated analytically for weak disorder, starting from the recursion relations for the  complex reflection- and transmission amplitudes.  The exact recursion relation for the reflection amplitudes is approximated by improved Born approximation forms which ensure that averaged reflection coefficients tend asymptotically to unity in the localized regime, for chain lengths $L=Na\rightarrow\infty$.
In contrast the familiar Born approximation of perturbation theory would not be adapted for the localized regime since it constrains the reflection coefficient to be less than one.  The proper behaviour of the reflection coefficient (and of other related reflection parameters) is responsible for various anomalies in the cumulants of $\ln g$, in particular for the well-known band center anomaly of the localization length.  While a simple improved Born approximation is sufficient for studying cumulants at a generic band energy, we find that a generalized improved Born approximation is necessary to account satisfactorily for numerical results for the band center anomaly in the mean of $\ln g$.  For the variance of $\ln g$ at the band center, we reveal the existence of a weak anomalous quadratic term proportional to $L^2$, besides the previously found anomaly in the linear term.  At a generic band energy the variance of $\ln g$ is found to be linear in $L$ and is given by twice the mean, up to higher order corrections which are calculated.  We also exhibit the $L=$independent offset terms in the variance, which strongly depend on reflection anomalies.

\end{abstract}


\maketitle

\section{Introduction}

The advent of the scaling theory of localization in $d$-dimensional
disordered systems
\cite{1} and more detailed developments of it in 1D \cite{2,3} and for
quasi 1D systems
\cite{4} has inaugurated a golden age for mesoscopic physics, particularly
the study of
transport phenomena.

The fundamental hypothesis in the scaling theory \cite{1} is that the
scaling of the logarithm
of a typical conductance $g$ as a function of a characteristic size $L$ of
the system is
described asymptotically for large $L$ by a universal function, $\beta(\ln
g)$, of a single parameter (SPS), namely $\ln g$
itself, such that $d\ln g/d\ln L=\beta(\ln g)$.  The function $\beta$
which may generally depend on dimensionality is
independent of $L$ and of microscopic  parameters in the system.  We recall
that in the studies of scaling in 1D systems
\cite{2,3} the  parameter $\ln (1+\rho)$, with $\rho=\frac{1}{ g}$ the
resistance, was identified as the
convenient scaling variable both in the localized regime ($\rho>>1$) where
it is self-averaging
and in the low resistance ($\rho<<1$) (quasi-metallic) regime.  This
variable reduces to $-\ln
g$ above for $\rho>>1$ (localized regime) and, thanks to the Landauer formula,
$\rho=\frac{r_L}{ t_L}$ (with $r_L$ and $t_L$ the reflection and
transmission coefficients of
the system, respectively), it coincides with $-\ln t_L\simeq -\ln g$.

In \cite{2,3} it was argued that the scaling theory of Abrahams et al.
\cite{1} had to be
interpreted in terms of the scaling of the distribution $P_g(g)$ of the
random conductance of
the system. SPS means that $P_g(g)$ is fully determined by a single
parameter such as e.g. the
mean logarithm, $<\ln g>$, which is itself defined by a scaling equation of
the above form
with $\ln g $ replaced by $<\ln g>$.

After years of debates the question of the validity of SPS in the theory of
Abrahams \textit{et al.}
\cite{1} remains still open and has recently been revived \cite{5,6,7,8,9}.
In particular, the
justification of the SPS hypothesis in the analyses \cite{2,3} rests on a
random phase
approximation (RPA) which assumes that the phases of the amplitude
reflection- and
transmission coefficients $R_L$ and $T_L$ (with $r_L=|R_L|^2, t_L=|T_L|^2$)
are uniformly
distributed over (0,2$\pi$) in the localization domain, i.e. for length
scales $L$ much larger
than the localization length $\xi$.  Despite the existence of strong
evidence, both numerical
\cite{10} and analytical \cite{11,12} for uniform phase distributions for
$L>>\xi$ in the 1D
Anderson model \cite{13} the SPS controversy is not resolved, essentially
because phases and
conductance are not independent random variables.

Doubts about the validity of results based on RPA have recently led Deych
and collaborators to reconsider the scaling problem
for the exactly soluble Lloyd model \cite{5} as well as to present simulation
results for the conductance distribution in the 1D Anderson model in the
region of fluctuation
states \cite{8}.  On the other hand, Schomerus and Titov \cite{6,7} (see
also Roberts
\cite{14}) have discussed simulation results for the first four cumulants
of $\ln g$ for the
Anderson model for weak disorder, both at a generic band energy and at
special energies
(band center and band edges) where ordinary perturbation theory fails.
They also discussed
the cumulants using a Fokker-Planck approach for the joint distribution of
$\ln g$ and of the
transmission phase \cite{6,7}.  Their results support the validity of the
lognormal SPS form of the
conductance distribution, at a generic band energy (unlike results of
Roberts\cite{14}), while showing deviations from SPS at
the
special energies.

Detailed analytical studies of conductance cumulants based on
properties of symmetric groups defined from generalized transfer matrices,
and on analytic
continuation procedures, have been published earlier by Slevin and Pendry
\cite{15} and by Roberts \cite{14}.  However, the
validation of these approaches rests on support from
numerical simulations and studies of limiting cases\cite{16}.

In this paper we adopt a new direct approach, circumventing RPA, for
studying the conductance distribution (the distribution
of $-\ln g=-\ln t_L$) analytically in the localized regime, for weak
disorder.  An essential ingredient of our analysis is the
identification of a general type of anomalous (non-perturbative) effects in
various complex reflection amplitude moments and
reflection coefficient moments of a finite chain in the localized regime,
both for a generic band energy and at the band
center.  These reflection anomalies strongly influence the logarithmic
conductance cumulants.  In particular, the anomaly of
the second moment of complex reflection amplitudes at the band center is
found to be responsible of the well-known
Kappus-Wegner anomalies in the localization length \cite{17,18,19} and in
the variance of the logarithmic
conductance \cite{6}.

In our analysis we choose to carry out the perturbation theory to fourth
order in the disorder (i.e. the random site
energies  in the Anderson model).  Now, by definition we have, for large
$L,\;\langle-\ln g\rangle=\langle-\ln
t_L\rangle=2L/\xi$ (whith
$\langle\ldots\rangle$ denoting averaging over disorder), where the inverse
localization length $1/\xi$ is proportional to the variance of the site
energies,
$\bar{\varepsilon^2}=\langle\varepsilon^2_n\rangle$
($\langle\varepsilon_n\rangle=0$), for weak disorder\cite{20}.  It
follows therefore that in the localized regime ($L>>\xi$) we have
$L\;\bar{\varepsilon^2}>>1$.  Hence it follows that the
$n$-th moment if $-\ln t_L$ is of order $(L\;\bar{\varepsilon^2})^n$.  This
shows that the fourth order perturbation theory
may yield a correct description of the mean and the variance of $-\ln g$
only.  While the study of the third- and fourth
cumulants, respectively at sixth and eighth orders, is thus left for the
future, our results for the first two cumulants at
a generic energy already rule out single-parameter scaling if effects
beyond leading order ($L\bar{\varepsilon^2}$) are
retained.

In Sect. II.A we present the analysis leading to exact formal expressions
for the first and second moments of $-\ln g$ in
terms  of cumulated reflection amplitudes- and reflection coefficients
moments.  In II.B we define successively the improved
Born approximation and a generalized improved Born approximation form of
the exact recursion relation relating the random
reflection amplitudes of samples of length $n\; a$ and $(n-1) a$,
respectively.  These approximate recursion relations are
solved exactly to obtain explicit expressions for reflection amplitudes-
and reflection coefficients moments.  The generalized
improved Born approximation turns out to be vital for studying the band
center anomaly in the localization length, in
particular for achieving good agreement with the numerical results of
Kappus and Wegner and others\cite{6,14,16,17,18,19}.  In Sect.
III we discuss our detailed analytical results for the localization length
and for the variance, both at a generic energy
and at the band center.  Some final remarks follow in Sect. IV.

\section{SCALING IN THE ANDERSON MODEL FOR WEAK DISORDER}

\subsection{The mean and the variance of $-\ln t_L$}

The Schr\"{o}dinger equation for a chain of $N$ disordered sites $1\leq
m\leq N$ of spacing
$a=1 (L=N)$ is

\begin{equation}\label{eq1}
\varphi_{n+1}+\varphi_{n-1}+\varepsilon_n\varphi_n=E\varphi_n\quad ,
\end{equation}
where the site energies $\varepsilon_n$, in units of a constant hopping
rate are mutually independent
variables, uniformly distributed between $-\frac{W}{2}$ and $\frac{W}{ 2}$
 ($<\varepsilon^{2p+1}_m>=0,<\varepsilon^2_m>=W^2/12,\langle\varepsilon^4_m\rangle=W^4/80$,
etc).  The disordered chain is connected as usual to semi-infinite
non-disordered chains
($\varepsilon_m=0$) at both ends, with sites $m<1$, and $m>N$, respectively.

The distribution of the transmission coefficient $t_N=|T_N|^2$
(conductance) for an electron
incident from the right with wavenumber $-k$ (energy $E=2\cos k$) may be
obtained by
solving the general recursion relations which connect the complex
transmission- (reflection-)
amplitudes $T_n (R_n)$ of a chain of $n$ sites with the corresponding
amplitudes for a chain
with one less disordered site, of length $n-1$.  These relations derived
in\cite{12} are, respectively,

\begin{equation}\label{eq2}
T_n=\frac{e^{ik}T_{n-1}}{1-i\nu_n (1+e^{2ik}R_{n-1})}\quad ,
\end{equation}
\begin{equation}\label{eq3}
R_n=
\frac{e^{2ik}R_{n-1}+i\nu_n (1+e^{2ik}R_{n-1})}
{1-i\nu_n (1+e^{2ik}R_{n-1})}\quad ,
\end{equation}
with
\begin{equation}\label{eq4}
\nu_n=\frac{\varepsilon_n}{ 2\sin k}\quad .
\end{equation}
The fundamental unitarity property,

\begin{equation}\label{eq5}
|R_n|^2+|T_n|^2=1\quad ,
\end{equation}
follows quite generally from (\ref{eq2}-\ref{eq3}) by expressing $|T_n|^2$,
using (\ref{eq2}), and $|R_n|^2$, using (\ref{eq3}),
 after rewriting the latter as $R_n=-1+(1+e^{2ik}R_{n-1})
[1-i\nu_n (1+e^{2ik}R_{n-1})]^{-1}$.

From (\ref{eq2}) we obtain, with the boundary conditions $T_0=1$ and $R_0=0$,

\begin{equation}\label{eq6}
-\ln t_N=\sum^{N}_{n=1}\biggl(\ln[1-i\nu_n(1+e^{2ik}R_{n-1}]+c.c.\biggr)
\quad ,
\end{equation}
which, together with (\ref{eq3}), is our starting point for studying the
probability distribution of $-\ln t_n\simeq -\ln
g$ via  the calculation of its moments and the corresponding cumulants.  We
choose to restrict our analysis of the
moments
$m_j=\langle(-\ln t_N)^j\rangle,j=1,2,\ldots$, to effects up to 4th order
in the random site energies and so we expand
(\ref{eq6}) in the form

\begin{equation}\label{eq7}
-\ln t_N=\sum^N_{n=1}\sum^4_{p=1}
\biggl(\frac{(-1)^{p+1}}{p}[-i\nu_n(1+e^{2ik}R_{n-1})]^p+c.c.\biggr)
\quad.
\end{equation}
As discussed in Sect. I systematic expansion to 4th order permits the
explicit study of the first two moments only, in
other  words the determination of the inverse localization length

\begin{equation}\label{eq8}
\frac{1}{\xi}=\frac{m_1}{2N}, N\rightarrow\infty
\quad,
\end{equation}
and of the variance,
\begin{equation}\label{eq9}
\text{var}\;(-\ln t_N)=m_2-m^2_1, N\rightarrow\infty
\quad.
\end{equation}

After some calculations, using (\ref{eq7}), the moments to 4th order in the
explicitated $\nu_n$ are reduced to the
following expressions (with $\bar{\nu^p}=<\nu^p_n>$):

\begin{equation}\label{eq10}
m_1=(\bar{\nu^2}-\bar{\frac{\nu^4}{2}})N
+\bar{\nu^2}
\biggl[(e^{2ik}R^{(1)}_N+\frac{e^{4ik}}{2}R^{(2)}_N)+c.c.\biggr] \quad ,
\end{equation}

\begin{equation}\label{eq11}
\begin{split}
m_2=-\bar{\nu^2} (e^{4ik}R^{(2)}_N+c.c.)+N\bar{\nu^4}+N (N-1) (\bar{\nu^2})^2\\
+2 (\bar{\nu^2}-\bar{\nu^4}) Q^{(1)}_N +\frac{\bar{\nu^4}}{2} Q^{(2)}_N \quad ,
\end{split}
\end{equation}
where
\begin{equation}\label{eq12}
R^{(p)}_N=\sum^N_{n=1} <R^p_{n-1}>
\quad ,
\end{equation}

\begin{equation}\label{eq13}
Q^{(p)}_N= =\sum^N_{n=1}<|R_{n-1}|^{2p}> \quad .
\end{equation}
The factorization of averages in (\ref{eq10}-\ref{eq11}) results from the
fact that the random
amplitudes $R_{n-1}$ and $R^*_{n-1}$ are linear functionnals depending on
$\nu_1,\ldots\nu_{n-1}$ but
not on $\nu_n$, as shown by iterating (\ref{eq3}), with $R_0=0$.  We have
used the fact that
the odd moments of $\nu_n$ are zero, which shows e.g. that
$<\nu_m\nu_n^pR_{m-1}R^*_{n-1}>=0,m\neq n, p=1,2$.  Also, in
(\ref{eq10}-\ref{eq11})we have systematically
dropped all terms of order higher than 4, in particular terms of the form
$<\nu^2_m\nu^2_n(R_{m-1}R^*_{n-1})^p>,p=1,2,m\neq
n$, in  (\ref{eq11}) which are of orders $4+2p$, unlike the corresponding
terms
with $m=n$ which lead to lower order anomalous effects, as shown below.

The Eqs. (\ref{eq10}-\ref{eq11}) reduce the study of the mean and of the
variance of $-\ln t_N$ to the calculation of the
quantities $R^{(1)}_N, R^{(2)}_N $ and $ Q^{(1)}_N , Q^{(2)}_N $, which we
refer to as fictitious {\em cumulated reflection
amplitudes- and cumulated reflection coefficients moments} (sums over
chains having lengths equal to rational fractions
of $Na$), respectively.  These sums are dominated for $N\rightarrow\infty$
by terms linear in $N$, reflecting the fact that
the amplitudes $R_n$ for $n\rightarrow\infty$ are described by an invariant
(stationary) distribution\cite{21,22} which is independent
of the initial site where the iteration of (\ref{eq3}) was started.  The
above cumulated moments will be studied in
Sect. III, using the improved Born approximations discussed below.

\subsection{Improved Born approximations}

The Born approximation for the random reflection coefficient $|R_n|^2$ for
weak disorder is obtained by assuming $R_n$ to
be  typically proportional to $\nu_n$ and approximating (\ref{eq3}) by the
linear recursion relation
$R_N=e^{2ik}R_{n-1}+i\nu_n$, whose solution

\begin{equation}\label{eq14}
R_n
=i\sum^n_{m=1}e^{2ik(n-m)}\nu_m
\quad ,
\end{equation}
yields

\begin{equation}\label{eq15}
\langle |R_n|^2\rangle=n\bar{\nu^2}=\frac{2n}{\xi_0}
\quad ,
\end{equation}
using the familiar perturbation expression for the Anderson localization
length\cite{20}.  Since the absolute limit of
$\langle |R_N|^2\rangle$ for a chain of length $N$ is unity, it follows
from (\ref{eq15}) that the Born approximation is not
suited for discussing the strong localization (localized) regime, where
$N\bar{\nu^2} >>1$ or $ N>>\xi_0$.  The same conclusion
also follows when using (\ref{eq14}) to calculate the second moment of the
reflection coefficient which enters in the
definition of $Q_N^{(2)}$.  In this case, we obtain from (\ref{eq14}), $\langle
|R_n|^4\rangle=n\biggl[\frac{3}{2}(n-1)(\bar{\nu^2})^2+\bar{\nu^4}\biggr]$,
which is also meaningless outside the perturbative domain
$(n\bar{\nu^2})^2<<1$.

In contrast, the study of the localized regime is possible if one uses the
improved (first) Born approximation where
(\ref{eq3}) is approximated, for weak disorder, by

\begin{equation}\label{eq16}
R_n=\frac{e^{2ik}R_{n-1}+i\nu_n}{1-i\nu_n}+\mathcal{O}(\nu^2_n)
\quad .
\end{equation}
For example, by iterating the recursion relation for the averaged
reflection coefficient, $\langle |R_n|^2\rangle$, obtained
from (\ref{eq16}) (using the fact that $R_{n-1}$, is independent of
$\nu_n$) and summing the resulting geometric series one
finds

\begin{equation}\label{eq17}
\langle |R_n|^2\rangle =1-(1-a_1)^n, a_1=\langle
\nu^2_n(1+\nu^2_n)^{-1}\rangle
\quad ,
\end{equation}
which has the desired limiting value of 1 for $n\;a_1>>1$, while reducing,
to leading order, to the perturbation result
(\ref{eq15}) in the opposite limit, $n\;a_1<<1$.  It follows that
(\ref{eq16}) represents the simplest approximation of Eq.
(\ref{eq3}) which permits a meaningful study of non-perturbative effects in
the Anderson model in the localized regime.

On the other hand, from the above discussion it is clear that the
expression for the second moment of $R_n$ obtained from
(\ref{eq14}), namely $\langle R_n^2\rangle=-e^{4ikn}n\bar{\nu^2}$ is also
invalid in the localized regime.  In contrast, by
solving the recursion relation for $\langle R_n^2\rangle$ obtained from
(\ref{eq16}) and summing the corresponding geometric
series we get, for any $E$,

\begin{equation}\label{eq18}
\langle R^2_n\rangle=\frac{c_2}{e^{4ik}c_1-1}+\mathcal{O}(\nu^4),\;
c_{p+1}=\langle\frac{\nu^{2p}_n}{(1-i\nu_n)^2}\rangle,\;
p=0,1;\; n\bar{\nu^2} >>1
\quad ,
\end{equation}
neglecting exponentially small terms proportional to $exp (n\;\ln c_1)$.
Here and in the following $\mathcal{O}(\nu^p)$
refers to contributions from terms of $p$th order in the random site
energies.  Eq. (\ref{eq18}) shows that while to leading
order
$\langle R^2_n\rangle$ is proportional to $\bar{\nu^2}$
 for $E\neq 0$ (as in perturbation theory), it is strongly enhanced by
non-perturbative effects, leading to
$\langle R^2_n\rangle =-\frac{1}{3}+\mathcal{O}(\nu^2)$ at $E=0$.  Using
this value one obtains from
(\ref{eq10}) and (\ref{eq12}) (with $\langle R^{(1)}_n\rangle
=\mathcal{O}(\nu^2)$) $\frac{1}{\xi}=\frac{\bar{\nu^2}}{3}$,
which corresponds to a reduction of 33\% of the result
$\frac{1}{\xi_0}=\frac{\bar{\nu^2}}{2}$ obtained by perturbation
theory\cite{20}.  This modification of the inverse localization length at
the band center has the same origin as the
well-known Kappus-Wegner\cite{17} anomaly.  But, clearly, its magnitude
calculated within the improved first Born
approximation is much too large since the effect obtained by Kappus and
Wegner and by others\cite{14,16,18,19}, using
various sophisticated approaches, range between 8 and 9\% of the
perturbation result.  This leads us to suggest a
generalised improved Born approximation of Eq. (\ref{eq3}) for dealing
specifically with the Kappus-Wegner anomaly, which
affects both $m_1$, and
$m_2$ at the band center, as shown by (\ref{eq10}) and (\ref{eq11}).  We
refer to this more accurate procedure (for the
band center) as the generalized improved Born approximation.  In this
approximation, besides the terms of the improved
first Born approximation, we retain the term $i\nu_n e^{2ik}R_{n-1}$ in the
numerator of (\ref{eq3}) as well as a further
term proportional to $R_{n-1}$ obtained by expanding the denominator around
the improved form $(1-i\nu_n)^{-1}$.  This
yields the approximate recursion relation

\begin{equation}\label{eq19}
R_n=e^{2ik}g_n R_{n-1}+f_n\; , \;
f_n=\frac{i\nu_n}{1-i\nu_n}\; , \; g_n=\frac{1}{(1-i\nu_n)^2}
\quad ,
\end{equation}
where we have ignored a third order term proportional to $R^2_{n-1}$.  In
Sect. 3.2 we return to a crude estimate of the
effects of this non-linear term to show that it does not affect the inverse
localization length $\frac{1}{\xi}$, nor the
variance of $-\ln t_N$, in the localized regime, to leading order in the
disorder.

\section{LOCALIZATION LENGTH AND VARIANCE OF $-\ln t_N$}

We first analyse the moments of $-\ln t_N$ at a generic band energy and
then we discuss how the calculations are modified to
account for the band center anomalies.

\subsection{Generic band energy}

By averaging (\ref{eq16}) over disorder (using the fact that $\nu_n$ is
independent of
$R_{n-1}$) and iterating the resulting recursion relation for $\langle
R_n\rangle$ in terms of a geometric series, which is
readily summed, we get

\begin{equation}\label{eq20}
R^{(1)}_N=\bar{\nu^2}u(N+u)+\mathcal{O}(\nu^4),u=(e^{2ik}-1)^{-1}
\quad ,
\end{equation}
where we have ignored exponentially small terms proportional to
$e^{-N\bar{\nu^2}}$ for weak disorder. Proceeding in a
similar  way with the equation  for $\langle R^2_n\rangle$ obtained from
(\ref{eq16}), we find

\begin{equation}\label{eq21}
R^{(2)}_N=\bar{\nu^2}v(N+v)+\mathcal{O}(\nu^4),v=(e^{4ik}-1)^{-1}
\quad .
\end{equation}
From (\ref{eq10}) and (\ref{eq20},\ref{eq21}) we then obtain

\begin{equation}\label{eq22}
m_1\equiv \langle-\ln t_N\rangle
=\biggl(\bar{\nu^2}+\frac{1}{2}[3(\bar{\nu^2})^2-\bar{\nu^4}]\biggr)
N-(\bar{\nu^2})^2 (2|u|^2+|v|^2)
\quad ,
\end{equation}
where the term linear in $N$ yields the familiar fourth order perturbation
expression for the
inverse localization length ($1/\xi$) at a generic energy \cite{14,22} and
the second term represents a new constant offset of
$m_1$.

Next we obtain the asymptotic form of the cumulated reflection coefficient
moments
$Q^{(1)}_N$ and $Q^{(2)}_N$. By summing (\ref{eq17}) over the disordered
sites we find

\begin{equation}\label{eq23}
Q^{(1)}_N=N-a^{-1}_1, Na_1\rightarrow\infty
\quad ,
\end{equation}
up to exponentially small terms.  The term $\frac{1}{a_1}$ is the leading
deviation of $Q^{(1)}_N $ from the unitarity limit
$(N)$ in the localized regime, $Na_1\sim N\bar{\nu^2}=\frac{2N}{\xi_0}>>1$,
in the improved Born approximation.\newline
In order to determine $ Q^{(2)}_N $ we have to solve the two-point
recursion relation for $\langle |R_{n-1}|^4\rangle$
derived  from (\ref{eq16}), namely

\begin{equation}\label{eq24}
\langle |R_n|^4\rangle=b_0\langle |R_{n-1}|^4\rangle
+ b_2+4b_1\langle |R_{n-1}|^2\rangle
-b_1 (e^{4ik}\langle R^2_{n-1}\rangle +c.c.)
\quad ,
\end{equation}
to 4th order non-vanishing terms.  Here we have defined

\begin{equation}\label{eq25}
b_p=\langle \nu^{2p}_n (1+\nu^2_n)^{-2}\rangle , p=0,1,2,\ldots
\quad ,
\end{equation}
and $\langle |R_{n-1}|^2\rangle $ is given by (\ref{eq17}) and $\langle
R_{n-1}^2\rangle =(\nu^2_n(1-i\nu_n)^{-2})v(1-e^{4ik(n-1)})+\mathcal{O}
(\nu^4)$.  As an example of the form of the
solution obtained by iterating such a recursion equation we refer, for
brievity's sake, to a  similar equation which is
solved in Scet. III.B.  In the 4th order expression (\ref{eq11}) we require
$Q^{(2)}_N$ to negative orders in $\nu$ up to zeroth order only.  After
performing successively the summation over sites in
the solution of (\ref{eq24}) and the further summation over sites in the
definition (\ref{eq13}) of $Q^{(2)}_N$ we obtain

\begin{equation}\label{eq26}
Q_N^{(2)}=\frac{1}{2b_1+b_2}\biggl[(4b_1+b_2)\biggl(N-\frac{1}{2b_1+b_2}\biggr)+
4\biggr]-\frac{4}{a_1}
\quad ,
\end{equation}

From (\ref{eq9}), (\ref{eq11}), (\ref{eq21}-\ref{eq23}) and
(\ref{eq25}-\ref{eq26}) we then obtain, to order $\nu^4$,

\begin{equation}\label{eq27}
\text{var}\; (-\ln t_N)=2\bar{\nu^2}(1-\bar{\nu^2})N-\frac{2\bar{\nu^2}}{a_1}
+\frac{\bar{\nu^4}(4b_1+3b_2)}{2(2b_1+b_2)^2}+\frac{(\bar{\nu^2})^2}{1-\cos 4k}
\quad ,
\end{equation}
which again involves a dominant term proportional to $N$ and a higher
constant offset term.
The leading term $2\bar{\nu^2} N$ in $\text{var}\;(-\ln t_N)$ coincides
with the result of various earlier theories\cite{2,6,14,15} for
a generic band energy.  We recall that its proportionality to the dominant
term of $\langle-\ln t_N\rangle$ in (\ref{eq22})
ensures single-parameter scaling of the lognormal conductance distribution
to lowest order, assuming that the higher cumulants
are negligible.  Our analysis reveals that this basic property is, in fact,
a direct consequence of the unitarity limit of the
reflection coefficient (\ref{eq17}) (which defines $Q^{(1)}_N $ for
asymptotic lengths in the localized regime).\newline
On the other hand, the cumulated reflection coefficients moments
(\ref{eq23}) and (\ref{eq26}) are responsible for the
existence of non-perturbative constant offset anomalies in the variance
(\ref{eq27}) in the localized regime, for a generic
energy.  Finally, we observe that Eqs. (\ref{eq22}) and (\ref{eq27})
involving terms proportional to $N$ and additional
constant offset terms conform to the ansatz of large-deviations statistics
for cumulants\cite{7,23}.  To zeroth order, the
offset in the variance (\ref{eq27}) reduces to the numerical constant $-2$.
An analogous zeroth order offset constant of
value $-\frac{\pi^2}{3}$ has been obtained in\cite{14}.

\subsection{Band center}

At the band center the denominators in (\ref{eq21}) are singular which
would require replacing them by their actual form
$e^{4ik}c_1-1$ in (\ref{eq18}) obtained from the improved Born
approximation.  However, as discussed in Sect. II.B we wish
to further improve the calculation of $R^{(2)}_N$ at the band center by
using the generalized improved Born
approximation (\ref{eq19}) of the exact recursion relation (\ref{eq3}).

By squaring (\ref{eq19}) and averaging over the disorder we obtain the
following relation for determining
$\langle R^2_n\rangle$:

\begin{equation}\label{eq28}
\langle R^2_n\rangle=e^{4ik}C\langle R^2_{n-1}\rangle+2 e^{2ik}D \langle
R_{n-1}\rangle-B
\quad ,
\end{equation}
where
\begin{equation}\label{eq29}
B=-\langle f^2_n\rangle, C=\langle g^2_n\rangle, D=\langle g_n f_n\rangle
\quad .
\end{equation}
On the other hand, by averaging (\ref{eq19}), solving for $\langle
R_n\rangle $ and performing the summation over sites in the
solution we get (with $f=\langle f_n\rangle, g= \langle g_n\rangle $)

\begin{equation}\label{eq30}
\langle R_n\rangle=\frac{e^{2ikn} g^n-1}{e^{2ik} g-1}f
\quad .
\end{equation}

By inserting (\ref{eq30}) in (\ref{eq28}), the recursion relation for
$\langle R^2_n\rangle $ takes the form

\begin{equation}\label{eq31}
\langle R^2_n\rangle= e^{4ik}C\langle R^2_{n-1}\rangle-F(e^{2ik}g)^{n-1}+F-B
\quad ,
\end{equation}
where
\begin{equation}\label{eq32}
F=2Df\frac{e^{2ik}}{1-e^{2ik}g}
\quad .
\end{equation}
The exact solution of (\ref{eq31}), with $R_0=0$ is given by

\begin{equation}\label{eq33}
\langle R^2_n\rangle=-B\; G^{n-1}-F\sum^{n-1}_{m=1}G^{n-m-1}(e^{2ik}g)^m+(F-B)
\sum^{n}_{m=2}G^{n-m}, G=e^{4ik} C
\quad ,
\end{equation}
which, after performing the geometric sums, reduces to

\begin{align}\label{eq34}
\langle R^2_n\rangle &= -G^{n-1} \biggl(B+ \frac{Fg}{e^{-2ik}G-g}
+\frac{B-F}{G-1}\biggr)\nonumber\\
&+ (e^{2ik}g)^{n-1}
\frac{Fg}{e^{-2ik}G-g}
+\frac{B-F}{G-1}
\quad .
\end{align}
Finally, we evaluate $ R^{(2)}_N $ defined by (\ref{eq12}) and
(\ref{eq34}), ingoring exponentially small terms proportional to
$e^{-N\bar{\nu^2}}$ (for weak disorder) in the localized regime.
Specializing to the band center $(k=\frac{\pi}{2})$ we have

\begin{equation}\label{eq35}
R^{(2)}_N=-\frac{1}{C(1-C)}\biggl(B-\frac{Fg}{C+g}-\frac{B-F}{1-C}\biggr)+\frac{
F}{(C+g)(1+g)}-\frac{N(B-F)}{1-C}
\quad ,
\end{equation}
where the anomalous denominator $1-C$ of order $\bar{\nu^2}$ is responsible
for the Kappus-Wegner correction in the localization
length and a corresponding anomaly in $\text{var}\;(-\ln t_N)$.  By
evaluating the quantities entering in (\ref{eq35}) and defined in
(\ref{eq19}), (\ref{eq29}) and (\ref{eq32}), for weak disorder, we obtain
explicitly

\begin{align}\label{eq36}
\bar{\nu^2} R^{(2)}_N &=-\frac{\bar{\nu^2} N}{10}
\biggl[1+\frac{\bar{\nu^4}}{2\bar{\nu^2}}+3\bar{\nu^2}+\mathcal{O}(\nu^4)\biggr]
\nonumber \\
&+\frac{1}{100}\biggl[1+3\bar{\nu^2}+\frac{4\bar{\nu^4}}{\bar{\nu^2}}+\mathcal{O
}(\nu^4)
\biggr]
\quad .
\end{align}
We recall that this quantity enters with opposite signs in
(\ref{eq10}) and (\ref{eq11}), respectively.

The final expression of $\langle-\ln t_N\rangle$ at the band center is then
obtained by substituting (\ref{eq20}) and
(\ref{eq36}) in (\ref{eq10}), which yields

\begin{align}\label{eq37}
\langle-\ln t_N\rangle
&=\frac{N}{10}\biggl[9\bar{\nu^2}+7(\bar{\nu^2})^2-\frac{11}{2}\bar{\nu^4}+\mathcal{O}(\nu^6)\biggr] \nonumber \\
&+
\frac{1}{100}\biggl[1+3\bar{\nu^2}+\frac{4\bar{\nu^4}}{\bar{\nu^2}}+\mathcal{O}(
\nu^4) \biggr]
\quad .
\end{align}
On the other hand, $\text{var}\;(-\ln t_N)$ is obtained from
(\ref{eq9}) and (\ref{eq11}) by inserting (\ref{eq23}) and
(\ref{eq26}), and (\ref{eq36}-\ref{eq37}). This leads to

\begin{align}\label{eq38}
\text{var}\;(-\ln t_N)
&=\frac{19}{100}(\bar{\nu^2})^2N^2+\biggl[2.182
\bar{\nu^2}-1.02(\bar{\nu^2})^2+0.01 \bar{\nu^4})
\biggr] N\nonumber \\
&-
\frac{1}{100}\biggl[2.01+6.06\bar{\nu^2}+8.08\frac{\bar{\nu^4}}{\bar{\nu^2}}+\mathcal{O}(\nu^4)
\biggr]
-\frac{2\bar{\nu^2}}{a_1}+\frac{\bar{\nu^4}}{2}\frac{(4b_1+3b_2)}{(2b_1+b_2)^2}
\quad .
\end{align}

The inverse localization length at $E=0$ obtained from (\ref{eq37}),

\begin{equation}\label{eq39}
\frac{1}{\xi}=\frac{1}{20}\biggl[9\bar{\nu^2}+7(\bar{\nu^2})^2-\frac{11}{2}\bar{
\nu^4}\biggr]
\quad ,
\end{equation}
may be compared to leading order with Thouless' perturbation expression
$1/\xi_0=\bar{\nu^2/2}$\cite{20}.  We thus find
that the  band center anomaly reduces the numerical coefficient of the
$\bar{\nu^2}$ term in $1/\xi$ by $10\%$ with respect
to its value in Thouless' expression.  For comparison, the earlier studies
of the band center anomaly in the inverse
localization length\cite{14,16,17,18,19} have yielded reductions of the
perturbation result ranging between 7.7 and 8.6
$\%$.  In view of the simplicity of our analytical treatment leading to a
simple transparent picture of the band center
anomaly in the localization length, we regard the agreement with the
numerical results of the earlier studies as rather
satisfactory.  Note also the existence of significant band center effects
in the coefficients of the quartic terms in
(\ref{eq39}), as shown by the comparison with the corresponding terms in
the perturbation result (Eqs. (\ref{eq8}) and
(\ref{eq22})) at a generic energy.

The comparison of the variance (\ref{eq38}) with the result (\ref{eq27})
for a generic energy reveals the existence of two
main types of non-perturbative band center anomalies: an enhancement of 9,1
\%  of the coefficient of the $\bar{\nu^2}N$
term, on the one hand, and the existence of an additional weak quadratic
term proportional to
$(N\bar{\nu^2})^2\sim(\langle-\ln t_N\rangle)^2$, on the other hand.  Such
a quadratic term at 4th order is obtained here
for the first time.  This term is comparable in magnitude with the leading
second order term for values of $N\bar{\nu^2}$
of the order of 10 or larger.  However, the addition of this term alone
does not spoil the single-parameter scaling
obtained when restricting to the second order in the disorder.  Finally it
appears that the accurate description of the
Kappus-Wegner anomaly in the localization length,  using the generalized
improved Born approximation, is crucial for
studying the variance since the magnitude of the new
$N^2$-term  in
(\ref{eq38}) is directly related to this anomaly.

We close this section with a brief remark about the effect of the non-linear
term

\begin{equation}\label{eq40}
\alpha R^2_{n-1}\equiv\frac{i\nu_n}{(1-i\nu_n)^2}e^{4ik}R^2_{n-1}
\quad ,
\end{equation}
which has been omitted on the r.h.s. of (\ref{eq19}) in the expansion of
(\ref{eq3}) for weak disorder.  For the purpose of a
crude estimate we approximate this term by the linearized form
$\alpha\langle R_{n-1}\rangle R_{n-1}$, where $\langle
R_{n-1}\rangle $ is given by (\ref{eq30}), and solve (\ref{eq19}) in the
presence of this approximate additional term.  Thus
we replace $g_n$ in (\ref{eq19})by
\begin{equation}\label{eq41}
g'_n=g_n\biggl(1-i\nu_n\frac{f}{1+g}\biggr)
\quad ,
\end{equation}
at the band center.  In this approximation the dominant effect of the
non-linear term in the moments (\ref{eq10}-\ref{eq11})
of $-\ln t_N$ arises via the denominator $1-C=1-\langle(g'_n)^2\rangle$ in
(\ref{eq35}) which is
responsible, in particular, for the Kappus-Wegner anomaly in the
localization length.  From the expansion of the parameters in
(\ref{eq41}) for weak disorder it follows that the correction term in this
expression leads to a 4th order correction
proportional to $(\bar{\nu^2})^2$ in $1-C$, thus leaving the dominant
second order term of $1-C$ unchanged.  Our estimatethus
shows that the non linear term (\ref{eq40}) has no effect on the mean and
on the variance of $-\ln t_N$ to leading order in
the disorder. \newline Finally, we note that a more realistic study of the
term (\ref{eq40}) might be to linearize it in
terms of the exact solution of the improved Born relation (\ref{eq16}) given by

\begin{equation}\label{eq42}
R_n=\sum^{n}_{m=1}e^{2ik(n-m)}\frac{i\nu_m}
{1-i\nu_m}\prod^{n}_{p=m+1}\frac{1}{1-i\nu_p}
\quad .
\end{equation}
However, the analytic solution of (\ref{eq19}) in the presence of such a
linearized form of (\ref{eq40}) is clearly very
complicated and will not be discussed.

\section{CONCLUDING REMARKS}

In this paper we have discussed a new approach for studying conductance
cumulants for weak disorder in the Anderson model in
the localized regime.  It is based on determining complex reflection
amplitudes using linear approximations of the exact non
linear recursion relation between the reflection amplitudes $R_n$ and
$R_{n-1}$ of disordered chains of lengths $na$ and
$(n-1)a$, respectively.  These approximate linear relations differ,
however, essentially from standard weak disorder
(perturbation) expansions in that they correctly account for the asymptotic
($n\rightarrow\infty$) unitarity property of the
averaged reflection coefficient $\langle |R_n|^2\rangle$, in the localized
regime, for any strength of a finite disorder.

Our analysis relates various anomalous effects of the disorder in the mean
and the variance of $-\ln t_N$ in the localized
regime to the above asymptotic behavior of the mean reflection coefficient
and/or to corresponding behavior of the second
moments of the reflection coefficient and of the complex reflection
amplitude (at the band center), respectively.  This
includes the well-known proportionality of
$\text{var}\; (-\ln t_N)$ to
$\langle-\ln t_N\rangle\sim\bar{\nu^2} N$ (to leading order) and the
existence of a leading numerically constant offset in
$\text{var}\; (-\ln t_N)$ at a generic energy.  It also includes both the
Kappus-Wegner\cite{17} anomaly and a leading constant offset
in $\langle-\ln t_N\rangle $ at the band center, as well as similar band
center anomalies in $\text{var}\;(-\ln t_N)$, in particular
the existence of a new term proportional to $N^2$ (which we have obtained
for the first time).

An important aspect of our treatment is that it does not rely on the
improper use of assumptions about phases such as the
phase randomization assumption which has frequently been invoked in
previous work\cite{2,3}.

The results of Sect. 3 indicate that beyond second order in the disorder
and ignoring the offset terms, the variance of
$-\ln t_N$ cannot be expressed in terms of the mean alone.  This rules out
single parameter scaling of the distribution of
$-\ln t_N$ even if one assumes the higher cumulants to be negligible (in
which case the distribution would be lognormal).  The
third, fourth $\ldots$ cumulants can, of course, also be studied using our
general approach, but, as shown in Sect. 1, this
requires perturbation expansions to 6th, 8th $\ldots$ order in the disorder.

Another application of the analysis of this paper would be the study of
conductance cumulants in coupled two- and three-chain
systems i.e. for few channel quasi-one dimensional systems.  Localization
in such systems has recently been discussed for weak
disorder both in the case where all the states at the fermi energy belong
to conducting bands of the pure few-channel
system and in the case where, on the contrary, the states for some of the
bands correspond to imaginary wavenumbers
(evanescent states) at the fermi energy\cite{24}.


\begin{thebibliography}{unsrt}
\bibitem{13} The invariant imbedding model studied in \cite{11} has
recently been shown \cite{12} to be equivalent to the
continuum limit of the tight-binding Anderson model for weak disorder.
\end{thebibliography}
\end{document}